\documentclass{ws-mpla}

\begin{document}

\markboth{V. Azcoiti, G. Di Carlo, A. Galante, V. Laliena}
{Phase Structure of Four Flavour QCD in the $T-\mu$ plane ...}

%
\catchline{}{}{}{}{}
%

\title{ PHASE STRUCTURE OF FOUR FLAVOUR QCD IN THE $T-\mu$\\
PLANE FROM A NEW METHOD FOR SIMULATIONS OF LATTICE\\ GAUGE THEORIES AT
NON-ZERO BARYON DENSITY}

\author{ VICENTE AZCOITI }

\address{Departamento de Fisica Teorica, Universidad de Zaragoza\\
Pedro Cerbuna 12, E-50009 Zaragoza, Spain}

\author{GIUSEPPE DI CARLO}

\address{INFN, Laboratori Nazionali del Gran Sasso 
I-67010 Assergi (L'Aquila), Italy}

\author{ANGELO GALANTE}

\address{Dipartimento di Fisica dell' Universita' di L'Aquila 
I-67100 L'Aquila, Italy and \\
INFN, Laboratori Nazionali del Gran Sasso 
I-67010 Assergi (L'Aquila), Italy}

\author{ VICTOR LALIENA}

\address{Departamento de Fisica Teorica, Universidad de Zaragoza\\
Pedro Cerbuna 12, E-50009 Zaragoza, Spain}

\maketitle


\begin{abstract}

We review a method for numerical simulations of lattice gauge theories
at non-zero baryonic chemical potential we recently proposed. We first report
on a test of the method using a solvable model and then 
present results for the phase structure of
four flavour QCD. For the first time the region of
chemical potential up to 1.4 $T_C$ is explored, finding a first order transition
line.

\end{abstract}


\section{Introduction}	

The existence and nature of deconfined phase(s) of QCD at high baryon
density is one of the open key issues of the theory of Strong Interaction
\cite{kogut}.
Progresses in this field are heavily delayed 
by the absence of a direct method of numerical
simulation based on an importance sampling Monte Carlo scheme.
In order to, at least partially, overcome this {\it empasse}, 
analytical approaches in the strong coupling limit using the 
Lagrangian \cite{monos} or Hamiltonian 
\cite{moslos1}, \cite{moslos2} approaches, as well as other indirect 
methods, have been proposed in the last years, which in principle allow
to extract physical informations of the behaviour of the vacuum of QCD
in the portion of the space parameter that is not accessible to
direct numerical simulations. Between the previously denoted 
indirect methods it is worthwhile to cite the Imaginary Chemical Potential 
approach and the Doubly Reweighting method 
\cite{imag,ph,mp,xiang,fodor,fodor2p1}.

Following this line we have recently proposed a different approach to the 
problem, which we hope can greatly help to improve the knowlegde 
in the field \cite{JHEP}. In this proceeding we present a concise
description of the method, some results of the test carried out in an exactly
solvable model (3-dimensional Gross-Neveu at non zero baryon density in
the large N limit) and finally discuss the new results we obtained 
concerning the phase structure of QCD with four flavour of light quarks 
at finite temperature and baryon density \cite{QCD4}.

\section{The Method}

Starting from the usual discretization of QCD with staggered fermions
with the chemical potential term introduced {\it a l\'a} Hasenfratz and
Karsch

\begin{eqnarray}
S &=& S_{\mathrm PG} + \frac{1}{2}\sum_n\sum^3_{i=1}
\bar\psi_n \eta_i (n) \left( U_{n,i}
\psi_{n+i} - U^\dagger_{n-i,i}\psi_{n-i}\right) \nonumber \\
&+& \frac{1}{2}\sum_{n} \bar\psi_n \eta_0 (n) \left( e^\mu  U_{n,0}\psi_{n+0} 
- e^{-\mu} U^\dagger_{n-0,0}\psi_{n-0}\right) + m \sum_n \bar\psi_n\psi_n
\label{hkaction}
\end{eqnarray}

we generalize the action introducing two independent parameters 
$x$ and $y$: the coefficient of the antihermitian and hermitian 
terms in the temporal part of the fermionic action. The usual action is
recovered when $x = \cosh (\mu a)$ and $y = \sinh (\mu a)$

\begin{eqnarray}
S &=& S_{\mathrm PG} +m \sum_n \bar\psi_n\psi_n +\frac{1}{2}\sum_n
\sum^3_{i=1}\bar\psi_n \eta_i (n)
\left( U_{n,i} \psi_{n+i} - U^\dagger_{n-i,i}\psi_{n-i}\right) \nonumber \\
&+& \frac{1}{2} x \sum_{n}\bar\psi_n \eta_0 (n)\left( U_{n,0}
\psi_{n+0} - U^\dagger_{n-0,0}\psi_{n-0}\right) \nonumber \\
&+& \frac{1}{2} y \sum_{n} \bar\psi_n \eta_0 (n)\left(  U_{n,0}
\psi_{n+0} + U^\dagger_{n-0,0}\psi_{n-0}\right)
\label{xyaction}
\end{eqnarray}

As in the case of (\ref{hkaction}) this action suffers, 
in general, from the sign problem; only when $x$ is real and
$y=i\bar y$ is imaginary the model defined through (\ref{xyaction}) can be
simulated with standard Monte Carlo methods.
Note that the case of imaginary chemical potential is a 
particular line in this subspace, defined by $x= \cos (\mu a)$ and 
$\bar y = \sin (\mu a)$. Studying the phase structure in the parameter 
space defined from
$\beta$ , $x$ and $\bar y$ and performing an analytical continuation
to real values of $y$ we can recover the phase structure of the
original model. 

In Ref. ~\refcite{JHEP} the reader can find an accurate discussion of what we
expect for the phase structure in the ($x,y$) and ($x,\bar y$) planes
and the motivations for such guess; 
it is of little interest, in our opinion, to repeat 
that discussion here, because we want to concentrate on the results 
concerning the critical line in the ($\mu,$T) plane. 
Assuming this attitude, let us sketch 
in some detail the procecdure used to extract the critical line 
in the cases under examination in the rest of the paper, {\it i.e.} 
Gross-Neveu model in three dimension and four flavours QCD.

Not to work in a three parameter space
we can look at the critical line in two dimensional sections; in
particular we can choose to work at fixed $\beta$, in the ($x$,$\bar y$) 
plane or at fixed $x$, in the ($\beta$,$\bar y$) plane.

In the first case we work at $\beta$ smaller than the critical value
of the zero density case and look for a critical line that starts 
from the $\bar y = 0$ axis moving towards larger values of $x$ and $\bar y$. 
This can be acomplished performing simulations at fixed $\bar y$ for several
values of $x$ starting from $\bar y=0$ and repeating the procedure
for several times at different values of $\bar y$.
Having determined the critical line in the ($x,\bar y$) plane we can
fit it with an even function of $\bar y$ and analytically continue
the line in the ($x,y$) plane. At this point the intersection of this
line with the $x^2-y^2=1$ line, that corresponds to real values of the
chemical potential, determines the critical value of $\mu a$ for the
given $\beta$. Repeating the whole
procedure for several values of $\beta$ will determine several points
on the critical line in the ($\beta,\mu a$) plane; values in physical
units can be obtained using the perturbative two-loop beta function.

The second alternative is simply described by exchanging the role of
$\beta$ and $x$. We work at fixed $x= \cosh(\mu a)$ 
and look for lines in the ($\beta,\bar y$) plane, analitycally 
continue to the ($\beta,y$) plane and intersect them with $y= \sinh (\mu a)$ 
to get the critical values in the ($\beta,\mu a$) plane, as before.


\section{Results in the Gross-Neveu model}

The Gross-Neveu model ({\it i.e.} four-Fermi model in 2+1 dimensions) 
is an ideal test bench for our approach: in the chiral limit 
it posseses a phase structure in the ($\mu$,T) plane that is similar 
to the one expected for four flavour QCD and it can be analytically 
solved in the large N limit; the exact solution allows us to determine
the critical line in the ($x,\bar y$) plane for a fixed $\beta$.
In Fig. 1 the whole procedure is illustrated (take in mind that the
variable in horizontal axis is $\bar y$ for the upper curve and $y$
for the two lower curves).

\begin{figure}[th]
\centerline{\psfig{file=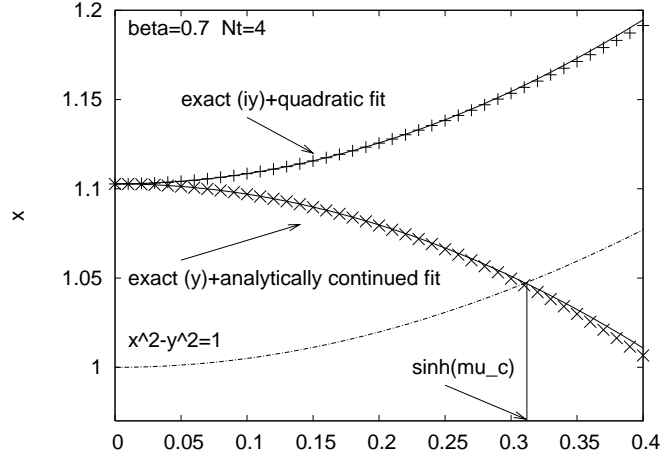,width=2.5in,angle=270}}
\vspace*{8pt}
\caption{Analytical continuation of the critical line from imaginary
to real $y$ and determination of the physical critical $\mu$; Gross-Neveu
model at $\beta=0.7$ and $N_t=4$.}
\end{figure}

Repeating this procedure for others values of $\beta$ produces the
critical structure of the model in the ($\mu,\beta$) plane 
depicted in Fig. 2. In this figure
the symbols stand for the critical points obtained with our method,
the continuous line represents the exact results for $N_t=4$ and the 
dashed line is what can be obtained in the same situation performing
an analytical extension starting from results obtained working in
the Imaginary Chemical Potential approach (in all cases the analytical
extrapolations rely on quadratic fits). It is evident a clear
advantage in using our new method for increasing chemical potential
values \cite{JHEP}.

\begin{figure}[th]
\centerline{\psfig{file=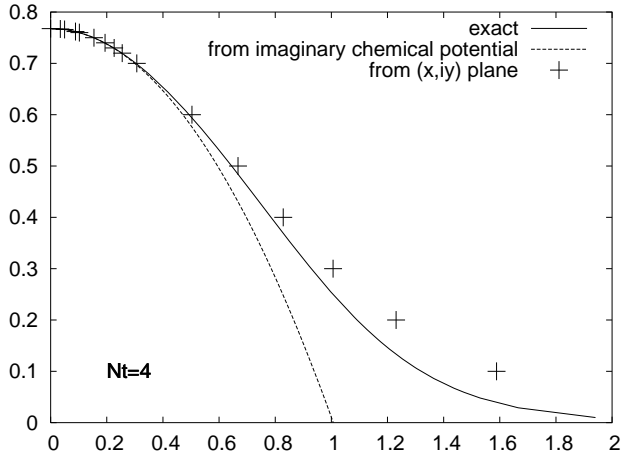,width=2.5in,angle=270}}
\vspace*{8pt}
\caption{Critical structure of Gross-Neveu model at $N_t=4$
in the $\mu,\beta$ plane.}
\end{figure}

\section{Four Flavour QCD}

While in the Gross-Neveu model we start from data (the critical points
at imaginary values of $y$) obtained 
solving seminumerically the gap equation, for QCD we have to rely
uniquely on Monte Carlo simulation. For this purpose we choose to 
work with a Hybrid Monte Carlo code for staggered fermions. 
The lattices we have used, due
to the fact that our method is quite expensive from a computational
point of view (for any point in the ($\mu$,T) plane we have to run the
Monte Carlo to span a two dimensional space), 
are limited to $8^3\times 4$ volumes at most. We choose a
quark mass of $ma=0.05$ in order to compare with existing results
obtained with Imagianry Chemical Poterntial and Doubly Reweighting 
methods \cite{mp,fodor}.

For each value of $x$, $\bar y$ and $\beta$ we run around $40000$
molecular dynamics trajectories of unit time, measuring plaquette,
chiral condensate and Polyakov loop for each trajectory. The simulations 
have been performed on Linux Clusters for a total of 100000 CPU hours.

We have worked, here, in the fixed $x$ scheme, choosing
5 values of $x$ corresponding to $\mu a$ of $0.2$, $0.25$, $0.3$, 
$0.4$ and $0.5$. For each $x$ value we perform a search of the critical
point in $\beta$ for a number of $\bar y$ values varying from 6 to 9,
using a Ferrenberg-Swendsen procedure to precisely determine the critical
values. In all cases the signal from the susceptibilities of the
three observables we used gave identical results for the extimation
of the critical points. The transition line was found to be a first
order one, with a latent heat essentially independent of the $\bar y$
value.

For the following stage we fitted the critical line in the 
($\bar y,\beta$)
plane, for each $x$, with a second order even polynomial:

\begin{equation}
\beta(\bar y) = \beta_0(x) + b_2(x) \bar y^2
\label{betafit}
\end{equation}

It is straightforward to continue the
critical line to real values of $y$ and intersect with the $y= \sinh(\mu a)$
vertical line to obtain the corresponding critical coupling
$\beta_c(\mu a)$. As the latent heat was constant along the
line in the ($\bar y,\beta$) plane, we assume his analytical extension
shares the same behaviour; hence we can affirm that
the transitions at different $\mu a$ are still first order, as
the one at $\mu=0$. This procedure is sketched in Fig. 3 for
a representative value of $x$ (as before the horizontal axis
is $\bar y$ for the upper curve and $y$ for the lower one).

\begin{figure}[th]
\centerline{\psfig{file=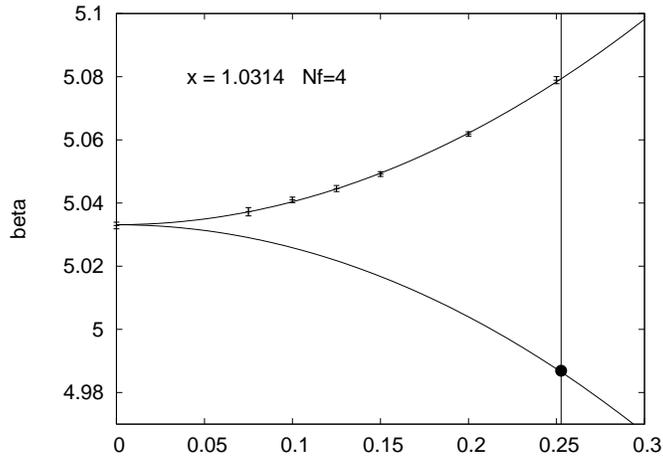,width=2.5in,angle=270}}
\vspace*{8pt}
\caption{Phase diagram at fixed $x$ in the ($\bar y,\beta$) plane,
its continuation in the ($y,\beta$) plane and determination of the
physical critical point.}
\end{figure}

Repeating this procedure for the five values
of $x$ (or $\mu a$) we have chosen we obtain the critical line
in the ($\mu a,\beta$) plane \cite{QCD4}. At this point we can rewrite our
results in physical units, as reported in Fig. 4, setting the
scale with the critical temperature at $\mu=0$ and using the two
loop beta function. In this figure there are also shown results obtained
by D'Elia and Lombardo \cite{mp} using the Imaginary Chemical 
Potential approach as well as a fit of out results with a power law:

\begin{equation}
\frac{T}{T_\mathrm{C}}\;=\;\left[1-c \,(\mu/T_\mathrm{C})^2\right]^p
\label{powerfit}
\end{equation}

Note that, if we take seriously this ansatz for the critical line
in the ($\mu/T_c,T/T_c$) plane, this leads to a prediction of the critical
chemical potential at zero temperature of $1.5 T_c \simeq 250$ MeV.

\begin{figure}[th]
\centerline{\psfig{file=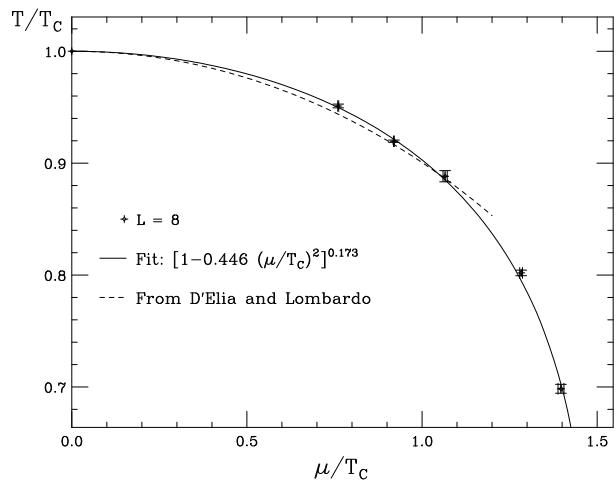,width=2.5in,angle=90}}
\vspace*{8pt}
\caption{Phase diagram in physical units.}
\end{figure}


\section*{Acknowledgments}

This work received financial support from CICyT (Spain),
project FPA2003-02948, from Ministerio de Ciencia y Tecnolog\'{\i}a 
(Spain), project BFM2003-08532-C03-01/FISI, and from
an INFN-CICyT collaboration. 
The authors thank the Consorzio Ricerca Gran Sasso that has provided
part of the computer resources needed for this work.
V.L. is a Ram\'on y Cajal fellow.



\end{document}